# Proposal for fiber optic data acquisition system for Baikal-GVD

V.A. Allakhverdyan[a], A.D. Avrorin[b], A.V. Avrorin[b], V.M. Aynutdinov[b],
R. Bannasch[c], Z. Bardačová[d], I.A. Belolaptikov[a], I.V. Borina[a], V.B. Brudanin[a],
N.M. Budnev[e], V.Y. Dik[a], G.V. Domogatsky[b], A.A. Doroshenko[b,1], R. Dvornický[a,d],
A.N. Dyachok[e], Zh.-A.M. Dzhilbkibaev[b], E. Eckerová[d], T.V. Elzhov[a], L. Fajt[f],
S.V. Fialkovsky[g], A.R. Gafarov[e], K.V. Golubkov[b], N.S. Gorshkov[a], T.I. Gress[e],
M.S. Katulin[a], K.G. Kebkal[c], O.G. Kebkal[c], E.V. Khramov[a], M.M. Kolbin[a],
K.V. Konischev[a], K.A. Kopański[h], A.V. Korobchenko[a], A.P. Koshechkin[b],
V.A. Kozhin[i], M.V. Kruglov[a], M.K. Kryukov[b], V.F. Kulepov[g], Pa. Malecki[h],
Y.M. Malyshkin[a], M.B. Milenin[b], R.R. Mirgazov[e], D.V. Naumov[a], V. Nazari[a],
W. Noga[h], D.P. Petukhov[b], E.N. Pliskovsky[a], M.I. Rozanov[j], V.D. Rushay[a],
E.V. Ryabov[e], G.B. Safronov[b], B.A. Shaybonov[a], M.D. Shelepov[b], F. Šimkovic[a,d,f],
A.E. Sirenko[a], A.V. Skurikhin[i], A.G. Solovjev[a], M.N. Sorokovikov[a], I. Štekl[f],
A.P. Stromakov[b], E.O. Sushenok[a], O.V. Suvorova[b], V.A. Tabolenko[e],
B.A. Tarashansky[e], Y.V. Yablokova[a], S.A. Yakovlev[c], and D.N. Zaborov[b]

[a] *1 Joint Institute for Nuclear Research, Dubna, Russia, 141980*

[b] *2 Institute for Nuclear Research, Russian Academy of Sciences, Moscow, Russia, 117312*

[c] *3 EvoLogics GmbH, Berlin, Germany, 13355*

[d] *4 Comenius University, Bratislava, Slovakia, 81499*

[e] *5 Irkutsk State University, Irkutsk, Russia, 664003*

[f] *6 Czech Technical University in Prague, Prague, Czech Republic, 16000*

[g] *7 Nizhny Novgorod State Technical University, Nizhny Novgorod, Russia, 603950*

[h] *8 Institute of Nuclear Physics of Polish Academy of Sciences (IFJ PAN), Krakow, Poland, 60179*

[i] *9 Skobeltsyn Institute of Nuclear Physics MSU, Moscow, Russia, 119991*

[j] *10 St. Petersburg State Marine Technical University, St. Petersburg, Russia, 190008*

*E-mail:* doroshenko@inr.ru

ABSTRACT: The first stage of the construction of the deep underwater neutrino telescope Baikal-GVD is planned to be completed in 2024. The second stage of the detector deployment is planned to be carried out using a data acquisition system based on fiber optic technologies, which will allow for an increased data throughput and more flexible trigger conditions. A dedicated test facility has been built and deployed at the Baikal-GVD site to test the new technological solutions. We present the principles of operation and results of tests of the new data acquisition system.

---

[1] Corresponding author.

## 1. Introduction

The large-scale neutrino telescope Baikal-GVD [1, 2] is under construction now. The beginning of the deployment of the telescope dates back to 2016, when the first cluster of the installation, which includes 288 photodetectors - optical modules (OM), was put into operation. By 2021, eight such clusters have been commissioned, and Baikal-GVD is currently the largest neutrino telescope in the Northern hemisphere. In the next three years, it is planned to increase the number of clusters to 14, covering a total effective volume of about 0.7 km$^3$ for the registration of astrophysical neutrinos. For the further expansion of the installation, the possibility of upgrading the data acquisition system based on a fiber-optic deep-water communication is being considered. This will increase the efficiency of the telescope operations by reducing the registration thresholds and organizing a more flexible trigger system. This article discusses the general approaches to modernization of the Baikal-GVD data acquisition system (DAQ) [3] with fiber-optic technologies and presents the first technical solutions, which are currently being tested in-situ as part of the experimental string of the neutrino telescope.

## 2. DAQ design

The design of the optical DAQ includes components associated with all basic elements of the detector, which are *sections*, *strings* and *clusters* of *optical modules* (OM). Each section consists of 12 OMs [4, 5]. Three sections located on one carrier cable form a string. Eight strings with an underwater data acquisition center are combined into a cluster. The design aims at providing an optic-fiber communication solution for the data stream, as well as trigger and sync signals. Another important requirement was the use of detachable optical connectors for underwater optical cables and minimization of their number, which is due to the technical conditions for the deployment of the strings from the ice of the lake.

For the DAQ implementation the CWDM technology (Coarse Wavelength Division Multiplexing) has been chosen, which is now widely used in the building of fiber-optic systems. CWDM multiplexers are passive devices that allow to organize up to 9 physical channels on a single fiber using the frequency division method. This approach makes it possible to transmit all information (the data stream, trigger, and sync) over a single optical fiber, which allows to limit each segment of the underwater network to one fiber-optic line.

## 3. In-situ tests

In order to test experimentally the possibilities of using CWDM technology to build the Baikal-GVD DAQ, an *experimental section* based on a fiber-optic data transmission system was installed in the telescope in 2020. The section included 12 optical modules (OM), two acoustic modems (AM) of the positioning system [6], and a section control module. To use all advantages of the high-speed optical transmission, the data acquisition unit of the section module (*Master*) was upgraded on the basis of FPGA Xilinx Zynq.

The experimental section was tested during 2020 in the mode of operation with two joint triggers: the coincidence of signals from neighboring optical modules (the *fast* trigger) and a jump-like increase in the pulse counting rate of the optical modules (the *slow* trigger). In the second case, the trigger was formed when the counting rate of any channel of the section exceeding during 1 ms the average rate by 4 standard deviations. This version of the trigger is focused on the selection of time-extended events, in particular, the registration of *slow monopoles* (the trajectory of the slow monopole at the intersection of the water volume should look like a "chain" of flashes resulting from the reactions of monopole catalysis of baryon decay).

During the year of the section operation, no malfunctions of the section operation were detected. The successful operation of the experimental section made it possible to start testing the experimental string, which is a full-scale prototype of the basic string of Baikal-GVD. In April 2021 the string was installed in Lake Baikal. The block diagram of the experimental string is shown in Figure 1

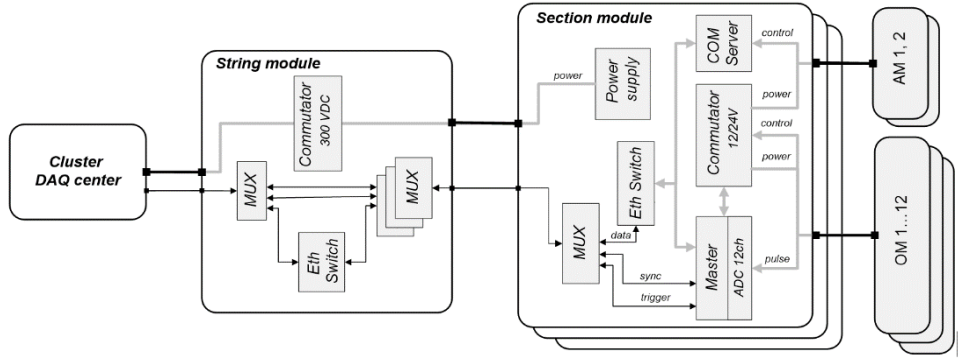

**Figure 1.** Block diagram of the experimental string.

The *experimental string* consists of three sections, a string control module and an optical cluster center. The construction of the sections is similar to the *experimental section*. The control modules of the sections are each connected to the string control module by two separate cables: an optical one with a single fiber and an electric one for the power supply. The fiber optic cables are equipped with underwater optical connectors. The signals received from the OMs are processed in the *Master* unit, which is equipped with a 12-channel ADC with a sampling rate of 200 MHz. The *Master* forms two optical channels (*trigger* and *sync*) and one electrical channel (Gigabit Ethernet), which is converted to optical using an Ethernet switch. The OMs and AMs of the section are controlled via a 16-channel power switch and a COM Server (MOXA NPort 5250). The input voltage of 300 VDC is supplied to the section control module via two independent lines: one is used to control electronics power supply, the second one provides power for OMs and AMs. In the string module, the data channels of the three sections are combined into one on the Ethernet switch. Therefore, each of the three sections in a string have a total of seven physical channels, one for data, three for trigger, and three for synchronization message. The data transmission via these channels to the cluster DAQ center is carried out via a CWDM multiplexer over a single optical fiber.

The *experimental string* started operation on 8 April 2021 in the mode of operation of two joint triggers, similar to the *experimental section* of 2020. The *fast trigger* initiated reading of information from all sections of the string, while the *slow trigger* operated only for the section that formed it. Currently, the main task of investigations with the experimental string is to evaluate the efficiency and reliability of the operation of the fiber-optic DAQ: both electronic components and underwater cable network.

There are two types of fiber-optic (FO) cable lines used for the experimental string. Most of the connections are made by commercial cable assemblies manufactured by DWTEK Co., Ltd, Taiwan. The Metal Shell Single Fiber Optic Bulkhead Connector Receptacles MSS-OP-BCR were used in combination with the Metal Shell Single Fiber Optic Cable Connector Plugs MSS-OP-CCP, installed on a radially sealed underwater FO cable manufactured by the same company (DWTEK MO1-I01590/OPY402 900um). Experimental string comprises of 5 such cable lines with different lengths from 3 to 750 m. To install the BCR in the openings of the glass sphere of the underwater modules, special stainless adapters were used to securely fix the connector and reduce the local pressure on the glass surface. One of the three sections is connected to the string using the experimental development of the Russian enterprise NPP "Starlink". The main condition for reliable optical communication is a low level and long-term stability of signal attenuation in fiber optic lines. To monitor this parameter, the power measurement function of transmitters and receivers built into the SFP modules was used. The threshold power of the NS-SFP 1.25 G CWDM optical transceivers used in the experimental string is -23 dBm (a signal with a power of 1 mW corresponds to 0 dBm). Figure 2 shows the time dependence of power of transmitters and receivers (wavelengths are 1510 nm, 1530 nm, 1550 nm, and 1570 nm) of the SFP modules of the *trigger* and *synchronization* channels for one sections of the string during about three months of operation. The power attenuation in these channels is about 10 dB, of which 2-3 dB is initially contributed by the fiber-optic connectors and cable line, and ~7 dB is due to the signal attenuation in the multiplexer. The signal power exceeds the threshold value of the receivers by more than 10 dB. At the same time, there are fluctuations in the power of the received signals, and the study of the causes of the increasing loss is one of the priority tasks of the set-up testing.

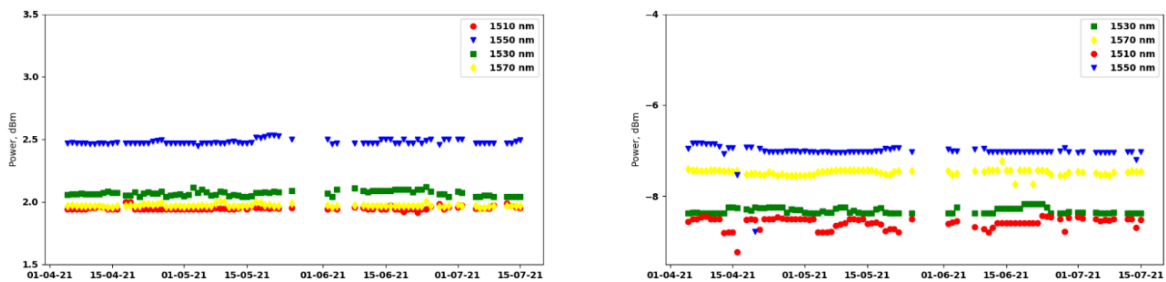

**Figure 2.** Time dependence of the power of the transmitters (left) and receivers (right) of optical transceivers for one section.

**4. Conclusion**

To improve the efficiency of Baikal-GVD and expand the possibility of reconfiguring its measurement system, research is being conducted on the possibilities of upgrading the data acquisition system on the basis of fiber optic communication. The first successful experience of such a system was obtained in 2020, when the experimental section with the upgraded DAQ was put into operation. In 2021, the research was continued and the experimental string, which includes three sections (36 OMs), was installed in Lake Baikal. In general, it is already possible to make a conclusion about the prospects of implementing a fiber-optic communication system based on CWDM technology at the Baikal-GVD. However, the problems associated with providing the installation with reliable deep-water optical cables have not been completely solved. In 2022, it is planned to continue research in this direction on the basis of additional experimental strings, which are planned to be installed in Lake Baikal.

This work was supported by the Ministry of Science and Higher Education of Russian Federation within the financing program of large scientific projects of the "Science" National Project (grant no. 075-15-2020-778).